\documentclass[twocolumn,showpacs,amsmath,prl,aps,amssymb,longbibliography,superscriptaddress]{revtex4-2}
\usepackage{graphicx,braket,float}
\usepackage{amsmath}
\usepackage{amsfonts}
\usepackage{amsbsy}
\usepackage{xcolor}
\usepackage{soul}
\usepackage{bm}
\usepackage[normalem]{ulem}
\usepackage[unicode]{hyperref}
\hypersetup{
    unicode=true, % non-Latin characters in Acrobat?s bookmarks
    plainpages=false,
    colorlinks=true,% false: boxed links; true: colored links
    linkcolor=blue,% color of internal links
    citecolor=blue,% color of links to bibliography
    filecolor=blue,% color of file links
    urlcolor=blue% color of external links
}
\usepackage{physics}
\usepackage{bbold}
\usepackage{epstopdf}

\begin{document}

\title{Berezinskii-Kosterlitz-Thouless transitions in an easy-plane ferromagnetic superfluid}

\author{Andrew P. C. Underwood}
\affiliation{Department of Physics, Centre for Quantum Science, and Dodd-Walls Centre for Photonic and Quantum Technologies, University of Otago, Dunedin, New Zealand}

\author{Andrew J. Groszek}
\affiliation{ARC Centre of Excellence for Engineered Quantum Systems, School of Mathematics and Physics, University of Queensland, Saint Lucia QLD 4072, Australia}
\affiliation{ARC Centre of Excellence in Future Low-Energy Electronics Technologies, School of Mathematics and Physics, University of Queensland, Saint Lucia QLD 4072, Australia}

\author{Xiaoquan Yu}
\affiliation{Graduate School of China Academy of Engineering Physics, Beijing 100193, China}
\affiliation{Department of Physics, Centre for Quantum Science, and Dodd-Walls Centre for Photonic and Quantum Technologies, University of Otago, Dunedin, New Zealand}

\author{P. B. Blakie}
\affiliation{Department of Physics, Centre for Quantum Science, and Dodd-Walls Centre for Photonic and Quantum Technologies, University of Otago, Dunedin, New Zealand}

\author{L. A. Williamson}
\affiliation{ARC Centre of Excellence for Engineered Quantum Systems, School of Mathematics and Physics, University of Queensland, Saint Lucia QLD 4072, Australia}

\date{\today}

\begin{abstract}
A two-dimensional (2D) spin-1 Bose gas exhibits two Berezenskii-Kosterlitz-Thouless (BKT) transitions in the easy-plane ferromagnetic phase. The higher temperature transition is associated with superfluidity of the mass current determined predominantly by a single spin component. The lower temperature transition is associated with superfluidity of the axial spin current, quasi-long range order of the transverse spin density and binding of polar-core spin vortices (PCVs). Above the spin BKT temperature, the component circulations that make up each PCV spatially separate, suggesting possible deconfinement analogous to quark deconfinement in high energy physics. Intercomponent interactions give rise to superfluid drag between the spin components, which we calculate analytically at zero temperature. We present the mass/spin superfluid phase diagram as a function of quadratic Zeeman energy $q$. At $q=0$ the system is in an isotropic spin phase with $\mathrm{SO}(3)$ symmetry. Here the fluid response exhibits a system size dependence, suggesting the absence of a BKT transition. Despite this, for finite systems the decay of spin correlations changes from exponential to algebraic as the temperature is decreased.
\end{abstract}

\maketitle

Spinor Bose gases boast a plethora of spin phases, providing an ideal system for studying equilibrium and non-equilibrium properties of phase transitions~\cite{Kawaguchi2012R,StamperKurn2013a}. At zero temperature a ferromagnetic spin-1 condensate with quadratic Zeeman energy $0<q<q_0$ is in the easy-plane phase, with $q_0$ a quantum critical point~\cite{Stenger1998a,zhang2003}. This phase exhibits two broken continuous symmetries: a $\mathrm{U}(1)$ symmetry associated with global phase coherence and an $\mathrm{SO}(2)$ symmetry associated with transverse spin coherence~\cite{Murata2007a}. Quenching the system from the polar ($q>q_0$) to the easy-plane phase has revealed rich non-equilibrium dynamics as the system orders to the new ground state~\cite{Sadler2006a,Leslie2009a,Saito2005a,Barnett2011,Saito2007b,Lamacraft2007a,Damski2007a,Barnett2011,anquez2016,Williamson2016a,Williamson2016b,williamson2019,prufer2018}.

An even richer phase structure is possible at finite temperature due to the multitude of possible defects and superfluid currents~\cite{Mukerjee2006a,James2011a,kobayashi2019b,Natu2011a,Kawaguchi2012a,kisszabo2005,Qiang2003a,kobayashi2019,sonin2018}. In 2D, it is well known that true long-range order is prohibited~\cite{mermin1966,hohenberg1967} and that the onset of superfluidity is instead associated with a BKT transition~\cite{berezinskii1972,kosterlitz1973,nelson1977}. In an easy-plane ferromagnetic spin-1 Bose gas, the two continuous symmetries give rise to two distinct BKT temperatures~\cite{kobayashi2019}. Beyond the existence of these transitions, however, little is known about the finite temperature phases of this system. At $q=0$, the order parameter manifold changes from $\mathrm{U}(1)\times \mathrm{SO}(2)$ to $\mathrm{SO}(3)$~\cite{Ho1998a}. In general, the nature of superfluidity in $\mathrm{SO}(3)$ systems is not well understood~\cite{kawamura1984,kawamura2010,kawamura1993,wintel1994,wintel1995,kobayashi2019}.

%\bcomment{We have no reference to Kobayashi's other work: PRL {\bf 123} 075303 (2019), with linearly-coupled 2-component Bose gases, finding role for half-quantized vortices (decoupled) or one BKT transition by molecule-antimolecule pairs (coupled)}.

%In three dimensions, finite temperature perturbative expansions around the condensate phase (e.g. Hatree-Fock or first order self-consistent approaches) show that the transition to the transverse magnetised phase is shifted from the zero temperature transition~\cite{Phuc2011a,Kawaguchi2012a}. In two dimensions an analysis is more difficult: although it is possible to expand around a quasi-condensate phase~\cite{mora2003}, diverging fluctuations invalidate such a treatment close to the superfluid transition.

In this work we explore the finite temperature behaviour of a 2D spin-1 ferromagnetic Bose gas in the easy-plane phase. Results are obtained via sampling of the dynamical evolution of the system (c.f.\! Ref.~\cite{kobayashi2019}, which employs a Monte Carlo Wolff algorithm). We observe that the system first transitions to a mass-superfluid state and then, at a lower temperature, transitions to a spin-superfluid state, in agreement with Ref.~\cite{kobayashi2019}. The mass transition is predominantly determined by the superfluidity of the $m=0$ spin component (with $m\in\lbrace-1,0,1\rbrace$ the spin-1 magnetic sublevels). The spin transition is driven by the binding/unbinding of PCVs, which consist of spatially confined equal and opposite circulations in the $m=\pm 1$ components. Above the spin BKT temperature, the $m=\pm 1$ component circulations spatially separate, suggesting possible deconfinement analogous to a color plasma. We identify superfluid drag between spin components, arising from spin interactions~\cite{fil2004,fil2005,carlini2021}, which we calculate analytically at zero temperature. We determine the $(T,q)$ superfluid phase diagram for $0<q<q_0$. At $q=0$, the fluid response exhibits a system size dependence, suggesting the absence of a BKT transition~\cite{kobayashi2019}. Despite this, for finite-sized systems the decay of correlations of total spin still change from exponential to algebraic as the temperature is decreased. The decay exponent is close to $1/2$ at the cross-over, twice that of $U(1)$ systems. The recent observation of thermalization of a quasi-1D easy-plane spin-1 Bose gas~\cite{prufer2022} demonstrates that our results could be observed in current experiments.

%In this intermediate temperature phase, coherence between the spin sublevels diminishes, despite the confining nature of the spin-exchange energy, in analogy with a color plasma in quantum chromodynamics.
%We identify mass and spin superfluid BKT transitions by the momentum response of the mass and axial spin currents respectively~\cite{foster2010}.

\paragraph{Formalism.}

We use a simple-growth stochastic Gross-Pitaevskii model that couples a three-component spinor field $\Psi=(\psi_1,\psi_0,\psi_{-1})^\mathrm{T}$ to a grand canonical reservoir at chemical potential $\mu$ and temperature $T$~\cite{gardiner2002,gardiner2003,bradley2008,Bradley2014a,cfieldRev2008},
\begin{equation}\label{dPsi}
i\hbar\dd{\Psi} = \left(1-i\gamma\right)\left(\mathcal{L}\left\{\Psi\right\}-\mu\Psi\right)\dd{t}+i\hbar\dd{W}.
\end{equation}
The nonlinear operator $\mathcal{L}\left\{\Psi\right\}$ reads
\begin{equation}
\mathcal{L}\left\{\Psi\right\}\!=\!\left[-\frac{\hbar^{2}\nabla^{2}}{2M}\mathbb{1}+qf_{z}^{2}+g_nn\mathbb{1}+g_s\sum_{\alpha}F_{\alpha}f_{\alpha}\right]\Psi,
\end{equation}
and describes time evolution arising from the kinetic energy, quadratic Zeeman shift, density-dependent interactions (coupling constant $g_n>0$), and spin-dependent interactions (coupling constant $g_s<0$). Here  $n=\Psi^{\dagger}\mathbb{1}\Psi$ and $F_{\alpha} = \Psi^{\dagger}f_{\alpha}\Psi$ are the density and spin density respectively, with $\mathbb{1}$ the identity matrix in spin space, and $f_{\alpha}$ the spin-1 matrices ($\alpha\in\{x,y,z\}$). The dimensionless rate $\gamma$ determines how strongly the system couples to the reservoir and $\dd{W}=(\dd{w}_1,\dd{w}_0,\dd{w}_{-1})$ are Gaussian distributed complex noise terms with correlations $\expval{\dd{w_{m}^{*}(\mathbf{r})}\dd{w_{m'}(\mathbf{r}^\prime)}} = 2\gamma k_{B}T\delta(\mathbf{r}-\mathbf{r}^\prime)\delta_{m,m'}\dd{t}/\hbar$. Stationary solutions to Eq.~\eqref{dPsi} sample the grand canonical ensemble and are independent of $\gamma$~\cite{gardiner2003}. In a uniform system, $q_0=2|g_s|n\approx 2|g_s|\mu/g_n$.

We simulate a condensate with weak spin interactions $|g_s|=0.1g_n$ on an $N\times N$ grid with periodic boundary conditions. We use a plane wave basis cut off at the thermal energy $k_BT$~\cite{prokofev2001,prokofev2002}. With an adjustment to account for our use of a square grid, this gives a grid spacing $\Delta x=\sqrt{2\pi\hbar^{2}/Mk_{B}T}$. We take $\mu\approx 5q_0$ as a convenient energy scale with associated length unit $x_\mu=\hbar/\sqrt{M\mu}$. We express the temperature in terms of the dimensionless quantity $\mathcal{T}=Mg_n k_B T/\hbar^2\mu$, which captures the dependence of thermodynamic properties on both temperature and chemical potential~\cite{prokofev2001,prokofev2002}. Equilibrium states are obtained by evolving Eq.~\eqref{dPsi} until $t\approx 10^5\hbar/\mu$. We then evolve the equilibrium state, sampling at intervals of $10 \hbar / \mu$, to build up an ensemble of $\sim 10^4$ states from which thermal averages are calculated. The system size and hence thermalization time diverge as $\mathcal{T}$ decreases, therefore we restrict our analysis to $\mathcal{T}\ge 0.05$.% and density unit $n_u= \mu/g_n$. For the temperatures considered in this work, total density fluctuations are small. Hence $n\approx n_\mu$ and $q_0\approx 2|g_s|n_\mu$.

\paragraph{Spin and mass BKT transitions.}

The two continuous symmetries in the easy-plane phase give rise to two superfluid currents at low temperatures~\cite{kobayashi2019}. The first is a mass superfluid current arising from the global phase coherence. The second is an $F_z$ spin superfluid current arising from coherence of the transverse spin $\mathbf{F}_\perp=(F_x,F_y)$~\footnote{The correspondence between $F_z$ spin current and transverse spin coherence follows from noting that gradients of transverse spin angle gives rise to an $F_z$ current~\cite{Yukawa2012}, or more formally since $f_z$ is the generator of transverse spin rotations.}. The mass superfluid density can be determined from the response of the system to slowly moving boundary walls, via the current-current response tensor,
\begin{equation}\label{response}
\chi(\mathbf{k})=\frac{M}{k_BT}\int d^2\mathbf{r}e^{-i\mathbf{k}\cdot\mathbf{r}}\langle \mathbf{J}(\mathbf{0})\mathbf{J}(\mathbf{r})\rangle.
\end{equation}
Here $\mathbf{J}=(\hbar/M)\operatorname{Im}(\sum_m\psi_m^\dagger\nabla\psi_m)$ is the total mass current, and angle brackets denote a thermal average. In the long-wavelength limit, the longitudinal component $\chi^L(\mathbf{k})$ of the tensor $\chi(\mathbf{k})$ is affected by the total fluid response, while the transverse component $\chi^T(\mathbf{k})$ of $\chi(\mathbf{k})$ is only affected by the normal fluid response~\cite{foster2010,pollock1987,Baym1968}. The mass superfluid density is then $\rho_n=\lim_{k\rightarrow 0}[\chi^L(\mathbf{k})-\chi^T(\mathbf{k})]$ (see Appendix D of~\cite{foster2010} for details~\footnote{We absorb a factor of $k_BT$ into the definition of $\chi$}). We extend this procedure to determine the $F_z$ spin superfluid density $\rho_s$ by considering the response to a ``spin dependent'' moving boundary, where the $m=\pm 1$ boundaries move in opposite directions and the $m=0$ boundary is stationary (in experiments, this could be engineered via spin-dependent light fields~\cite{schmidt2016}). We then replace $\mathbf{J}$ by $\mathbf{J}_z=(\hbar/M)\operatorname{Im}(\sum_{m,m^\prime}\psi_m^\dagger (f_z)_{mm^\prime}\nabla\psi_{m^\prime})$ in Eq.~\eqref{response}.

%This boundary motion only affect the normal component $F_z^\prime$ of the $F_z$ spin density and hence the spin superfluidity $\rho_s$ can be obtained from $\rho_s=F_z-F_z^\prime$. For boundary motion along $x$ the normal fluid components can be determined from current-current correlations as,
%\begin{equation}\label{rhomu}
%\sigma^\prime=\frac{M}{k_BT}\int d^2\mathbf{r}\langle J_\sigma^x(\mathbf{0})J_\sigma^x(\mathbf{r})\rangle,\hspace{0.5cm}\sigma=n,F_z.
%\end{equation}
%Here $J_n^x=(\hbar/M)\operatorname{Im}(\sum_m\psi_m^\dagger\partial_x\psi_m)$ is the $x$-component of the total mass current and $J_{F_z}^x=(\hbar/M)\operatorname{Im}(\sum_{m,m^\prime}\psi_m^\dagger (f_z)_{mm^\prime}\partial_x\psi_m^\prime)$ is the $x$-component of the total $F_z$ spin current.}

%The ground state of a uniform system depends on $q/q_0$, with $q_0=2|c_s|n_0$ and $n_0$ the uniform density. For $0<q<q_0$, the ground state is magnetized in a plane perpendicular to the Zeeman field, giving rise to a non-zero transverse magnetization $\mathbf{F}_\perp=(F_x,F_y)$. This state has both gauge and spin symmetry, arising from energy the invariance of the energy under both global phase and transverse spin rotations. Hence the ground state manifold is $\mathrm{U}(1)\times \mathrm{SO}(2)$. The point $q_0$ is a quantum critical point separating the easy-plane phase from the polar phase (all occupation in the $m=0$ component), which possesses only a $\mathrm{U}(1)$ symmetry. 

\begin{figure}
\includegraphics[width=0.5\textwidth]{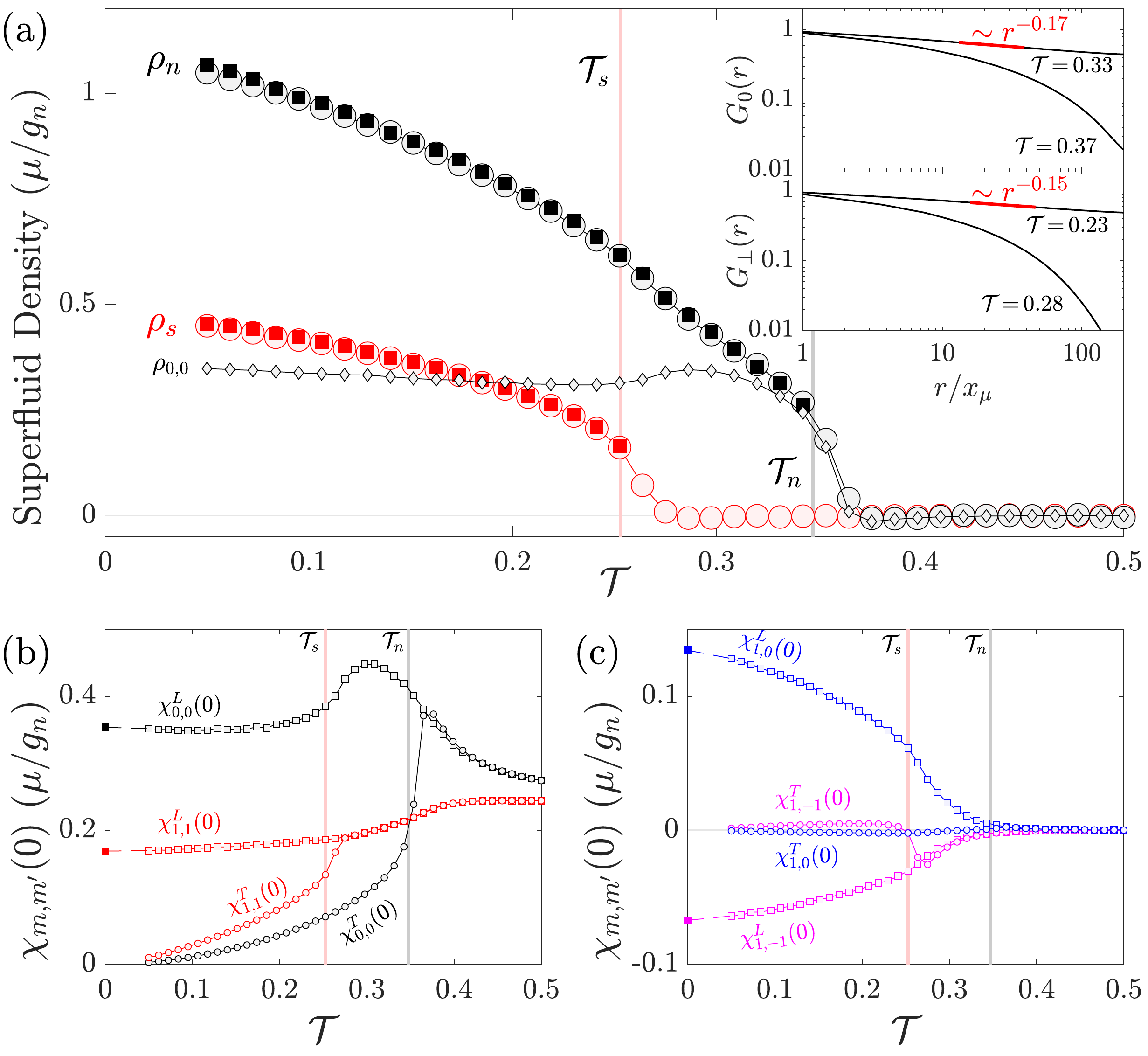}
\caption{\label{superfluid}(a) Mass (black circles) and $F_z$ spin (red circles) superfluid densities, determined from Eq.~\eqref{response}, as a function of dimensionless temperature $\mathcal{T}$. With reducing $\mathcal{T}$ the system first exhibits mass superfluidity and subsequently $F_z$ spin superfluidity. The superfluid densities obtained from fitted $\eta_{n,s}$, via $\rho_{n,s}=Mk_BT/2\pi\hbar^2\eta_{n,s}$, coincide with those determined from Eq.~\eqref{response} (matching black and red squares). We identify the precise BKT temperatures $\mathcal{T}_{n,s}$ to be the temperatures where $\eta_{n,s}=1/4$. Diamonds are $\rho_{0,0}$ (see text). Inset: correlations of $\psi_0$ decay algebraically below the mass superfluid transition ($G_0\sim r^{-\eta_n}$), while correlations of $\mathbf{F}_\perp$ decay algebraically below the spin superfluid transition ($G_\perp\sim r^{-\eta_s}$). Red lines are fits. (b), (c) Transverse (circles) and longitudinal (unfilled squares) components of $\lim_{k\rightarrow 0}\chi_{m,m^\prime}(\mathbf{k})$. Superfluid drag $\rho_{0,1}$ $(\rho_{-1,1})$ appears below $\mathcal{T}_n$ $(\mathcal{T}_s)$. The analytic zero-temperature $\chi_{m,m^\prime}^L(0)$ is indicated by filled squares. [Results for $q=0.03\mu\approx 0.15q_0$, $N=256$.]}
\end{figure}

In Fig.~\ref{superfluid} we plot the mass and spin superfluid densities determined from Eq.~\eqref{response}. Clearly evident are the two distinct BKT temperatures $\mathcal{T}_n$ and $\mathcal{T}_s$ associated with the onset of mass and spin superfluidity, respectively. Below the mass (spin) BKT temperature, two-point correlations of $\psi_0$ ($\mathbf{F}_\perp$) change from decaying exponentially to decaying algebraically,
\begin{equation}
\begin{split}
G_0(r)=\;&\frac{\langle\psi_0(\mathbf{0})^\dagger\psi_0(\mathbf{r})\rangle}{\langle \psi_0(\mathbf{0})^\dagger \psi_0(\mathbf{0})\rangle}\sim r^{-\eta_n}\hspace{0.5cm}(\mathcal{T}\le \mathcal{T}_n),\\
G_\perp(r)=\;&\frac{\langle \mathbf{F}_\perp(\mathbf{0})\cdot\mathbf{F}_\perp(\mathbf{r})\rangle}{\langle \mathbf{F}_\perp(\mathbf{0})\cdot\mathbf{F}_\perp(\mathbf{0})\rangle}\sim r^{-\eta_s}\hspace{0.5cm}(\mathcal{T}\le \mathcal{T}_s),
\end{split}
\end{equation}
see inset to Fig.~\ref{superfluid}(a). The mass (spin) superfluid densities estimated from the decay exponents $\eta_n$ ($\eta_s$), via $\rho_{n,s}=Mk_BT/2\pi\hbar^2\eta_{n,s}$~\cite{kosterlitz1973}, show excellent agreement with those determined from Eq.~\eqref{response}, see Fig.~\ref{superfluid}. Precisely at the respective BKT temperatures the exponents take on a universal value of $1/4$~\cite{nelson1977}; hence we identify $\mathcal{T}_n$ and $\mathcal{T}_s$ in Fig.~\ref{superfluid} as the temperatures where $\eta_{n,s}=1/4$.

\emph{Superfluid drag.} We can also examine the more general response functions
\begin{equation}\label{responsemm}
\chi_{m,m^\prime}(\mathbf{k})=\frac{M}{k_BT}\int d^2\mathbf{r}e^{-i\mathbf{k}\cdot\mathbf{r}}\langle \mathbf{J}_m(\mathbf{0})\mathbf{J}_{m^\prime}(\mathbf{r})\rangle,
\end{equation}
with $\mathbf{J}_m=(\hbar/M)\operatorname{Im}(\psi_m^\dagger\nabla\psi_m)$. Note $\chi_{m,m^\prime}=\chi_{m^\prime,m}$ and, due to symmetry under $m=1\leftrightarrow m=-1$, $\chi_{m,m^\prime}=\chi_{-m,-m^\prime}$. Defining $\rho_{m,m^\prime}\equiv \lim_{k\rightarrow 0}[\chi_{m,m^\prime}^L(\mathbf{k})-\chi_{m,m^\prime}^T(\mathbf{k})]$, with  $\chi_{m,m^\prime}^T$ ($\chi_{m,m^\prime}^L$) the transverse (longitudinal) components of $\chi_{m,m^\prime}$, the mass and spin superfluid densities can be decomposed as $\rho_{n}=\sum_{m,m^\prime}\rho_{m,m^\prime}$ and $\rho_{s}=\sum_{m,m^\prime}mm^\prime\rho_{m,m^\prime}$. We find that mass superfluidity is primarily determined from $\rho_{0,0}$ at the mass transition, see Fig.~\ref{superfluid}(a). The transverse and longitudinal components of the four independent $\chi_{m,m^\prime}$ are shown in Fig.~\ref{superfluid}(b) and Fig.~\ref{superfluid}(c). Off-diagonal contributions $\rho_{m,m^\prime\ne m}$ indicate ``superfluid drag'' between components $m$ and $m^\prime$~\cite{nespolo2017}, see Fig.~\ref{superfluid}(c). In other multicomponent systems, superfluid drag occurs via current-current coupling and is known as the Andreev-Bashkin effect~\cite{andreev1976,nespolo2017}, whereas in the spin-1 system it occurs due to inherent intercomponent interactions~\cite{fil2004,fil2005,carlini2021}.

The zero-temperature $\chi_{m,m^\prime}^L(0)$ can be obtained analytically as follows. Beginning with a stationary, uniform system, we impart current $(n_{m}\hbar k_{m}/M)\hat{\mathbf{x}}$ into spin component $m$ (with $n_{m}=|\psi_{m}|^{2}$). Due to intercomponent interactions, this will induce currents $(n_{m^\prime}\hbar k_{m^\prime}/M)\hat{\mathbf{x}}$ in components $m^\prime\neq m$. We obtain the $k_{m^\prime\neq m}/k_{m}$ by minimizing the additional kinetic energy $\delta E = (\hbar^{2}/2M)\sum_{m^\prime\neq m}n_{m^\prime}k_{m^\prime}^{2}$ subject to the constraint $k_{1}+k_{-1}-2k_{0}=0$ imposed by minimization of the spin-interaction energy. Defining a wavenumber $n_{m}k\equiv \sum_{m^\prime}n_{m^\prime}k_{m^\prime}$ and comparing with the relation $n_m=\sum_{m^\prime}\chi_{m,m^\prime}^L(0)$ we surmise $\chi_{m,m^\prime}^{L}(0)=n_{m^\prime}k_{m^\prime}/k$, where $k_{m^\prime}/k$ has an implicit dependence on $m$. This is confirmed in Fig.~\ref{superfluid}(b) and Fig.~\ref{superfluid}(c).

\paragraph{Topological properties and confinement.}

\begin{figure}
\includegraphics[width=0.49\textwidth]{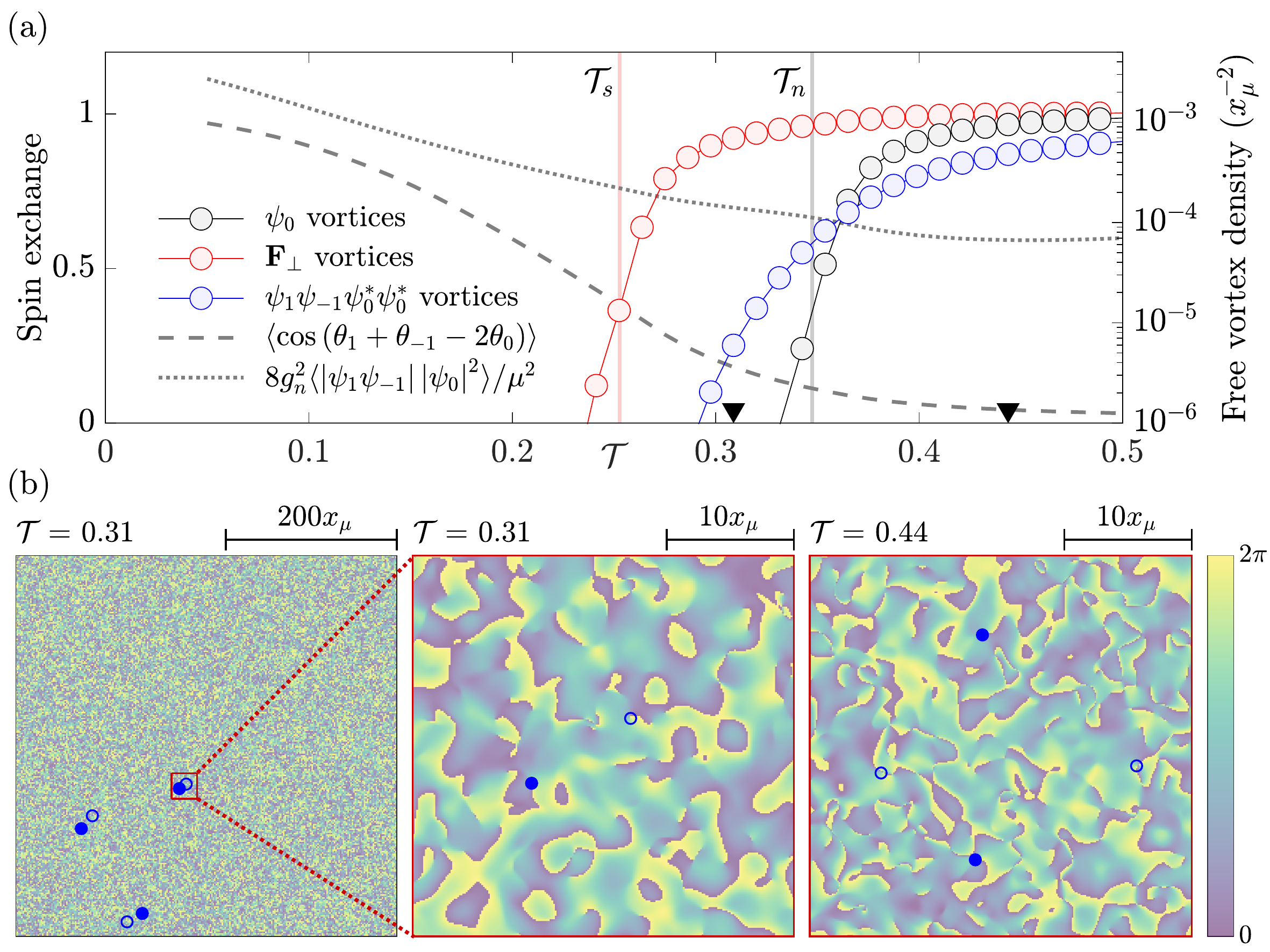}
\caption{\label{vortices}(a) Right axis: Free $\mathbf{F}_\perp$ vortices proliferate for $\mathcal{T}>\mathcal{T}_s$ (red circles), which can be identified as PCVs for $\mathcal{T}<\mathcal{T}_n$. Free $\psi_0$ vortices proliferate for $\mathcal{T}>\mathcal{T}_n$ (black circles). Free $\psi_1\psi_{-1}\psi_0^*\psi_0^*$ vortices proliferate for $\mathcal{T}\gtrsim 0.3$ (blue circles), suggesting deconfinement of PCVs, analogous to a color plasma. Left axis: coherence between $\psi_1\psi_0^*$ and $\psi_{-1}^*\psi_0$, measured by $\langle\cos(\theta_1+\theta_{-1}-2\theta_0)\rangle$ (gray dashed line), decreases with increasing temperature, along with a decrease of the ``coupling strength'' $\langle|\psi_1\psi_{-1}||\psi_0|^2\rangle$ (gray dotted line). (b) Profiles of $\theta_{1}+\theta_{-1}-2\theta_{0}$ at two temperatures (inverted triangles in (a)). Unfilled (filled) circles denote positive (negative) free $\psi_{1}\psi_{-1}\psi_{0}^{*}\psi_{0}^{*}$ circulations. [Results for $q=0.03\mu$, $N=256$.]} %\textcolor{red}{Match lines in fig 1 and 2 and check $\psi_1$ versus $\psi_1\psi_0^*$}}
\end{figure}

%\caption{\label{vortices}(a) Left axis: For $\mathcal{T}>\mathcal{T}_s$ there is a proliferation of free $\mathbf{F}_\perp$ vortices (red circles), and for $\mathcal{T}>\mathcal{T}_n$ there is a proliferation of free $\psi_0$ vortices (black circles). Right axis: Axial spin fluctuations increase with increasing $\mathcal{T}$, decreasing $|\psi_1\psi_{-1}||\psi_0|^2$ (gray dotted line). Accompanying this, $\cos(\theta_1+\theta_{-1}-2\theta_0)$ (gray dashed line) decreases, indicating a decoherence of the $\psi_{\pm 1}$ spin components. (b) Profiles of transverse spin density at different temperatures (inverted triangles in (a)) showing increasing spin disorder. Enhanced regions show free vortices. Note the presence of free $\psi_{\pm 1}$ vortices not clearly paired with a circulation of $\mathbf{F}_\perp$. Hence PCVs at this temperature are ``deconfined'', analogous to a color plasma. [Results for $q=0.03\mu\approx 0.15q_0$.]}

In the easy-plane phase the system supports two topologically distinct vortices, associated with the $\mathrm{U}(1)$ and $\mathrm{SO}(2)$ symmetries respectively. The destruction of mass superfluidity coincides with a proliferation of free $\psi_0$ vortices, see Fig.~\ref{vortices}(a), consistent with the finding that $\rho_n\approx \rho_{0,0}$ close to the mass BKT transition (Fig.~\ref{superfluid}). The destruction of spin superfluidity coincides with a proliferation of free $\mathbf{F}_\perp$ vortices, see Fig.~\ref{vortices}(a). Since $\psi_0$ is coherent at this temperature, these circulations can be identified as PCVs, rather than Mermin-Ho vortices~\cite{isoshima2001,khawaja2001,mizushima2002}. [Free vortices are identified by convolving the relevant field with a Gaussian filter before detecting divergences in the vorticity field. We choose a filter of width $5\sqrt{5}x_\mu$, which is on the order of the core size of a cold PCV~\cite{williamson2021}, with $\sqrt{5}x_\mu\approx \hbar/\sqrt{2M|g_s| n}$ the approximate spin healing length.] %Note that $\psi_0$ is coherent for $\mathcal{T}<\mathcal{T}_n$. Hence, the $\mathbf{F}_\perp$ vortices that destroy spin superfluid order can be identified as PCVs rather than Mermin-Ho vortices~\cite{isoshima2001,khawaja2001,mizushima2002}.

%\footnote{Free vortices are identified by first convolving the relevant field with a Gaussian filter with standard deviation $\sigma=8x_\mu$, which removes fluctuations smaller than $\sigma$, and then detecting divergences in the vorticity field.}

%\textcolor{red}{Remove: Note that in general, the $\psi_0$ current is not equal to the mass current, and in fact the mass current may not even be irrotational~\cite{Kawaguchi2012R,lamacraft2008}. However, if the spin-exchange energy is minimised, we can attribute mass current to flow of $\psi_0$, as the remaining contribution from the $m=\pm 1$ components cancel, and the mass current will be irrotational. The superfluid behaviour is consistent with this picture, which is also a requisite for assuming the order parameter manifold is $\mathrm{U}(1)\times \mathrm{SO}(2)$.}
%A PCV consists of equal and opposite phase winding of $\psi_\pm $ components with zero $\psi_0$ circulation~\cite{isoshima2001}, and is distinct from the Mermin-Ho spin texture~\cite{khawaja2001,mizushima2002}.
%Hence, as expected from the two superfluid phases in Fig.~\ref{superfluid}, our system has two corresponding topologically ordered phases: one with global phase topological order and no spin phase topological order, and a lower temperature phase with both global phase and spin phase topological order.

A single $\mathbf{F}_\perp$ vortex consists of equal and opposite phase windings of $\psi_1\psi_0^*$ and $\psi_{-1}\psi_0^*$ bound by the spin-exchange energy $2g_s\operatorname{Re}\psi_1\psi_{-1}\psi_0^*\psi_0^*$~\cite{isoshima2001,Turner2009}. This spin-exchange energy increases linearly with separation between the $\psi_{\pm 1}\psi_0^*$ vortices~\cite{Williamson2016c}, analogous to confinement in quantum chromodynamics~\cite{son2002,tylutki2016,eto2018confinement}. We find that the coherence between $\psi_1\psi_0^*$ and $\psi_{-1}^*\psi_0$, measured by $\langle\cos(\theta_1+\theta_{-1}-2\theta_0)\rangle$ (with $\theta_m$ the phase of $\psi_m$), decreases with increasing temperature, see Fig.~\ref{vortices}(a). Furthermore, there is a proliferation of free vortices in the quantity $\psi_1\psi_{-1}\psi_0^*\psi_0^*$ for $\mathcal{T}\gtrsim 0.3$, see Fig.~\ref{vortices}(a),(b), indicating spatial separation of $\psi_1\psi_0^*$ and $\psi_{-1}\psi_0^*$ vortices. This suggests possible PCV deconfinement, analogous to deconfinement in a color plasma. In such a plasma, this is enabled by a decrease in the strong-force coupling strength with increasing temperature~\cite{braun2007,pasechnik2017}. Analogously, the spin-exchange ``coupling strength'' $\langle|\psi_1\psi_{-1}||\psi_0|^2\rangle$ decreases with increasing temperature, see Fig.~\ref{vortices}(a). Ignoring fluctuations in $|\psi_0|$ and $n$, this coupling strength is $\propto\sqrt{n^2-F_z^2}$ and hence diminishes with increasing fluctuations of $F_z$ \footnote{Note that in the quadratic regime, fluctuations of $F_z$ increase with increasing transverse spin disorder, since both are affected by the same Bogoliubov mode~\cite{Symes2014a}.}.

\paragraph{Spin/mass superfluid phase diagram.}

\begin{figure}
\includegraphics[width=0.5\textwidth]{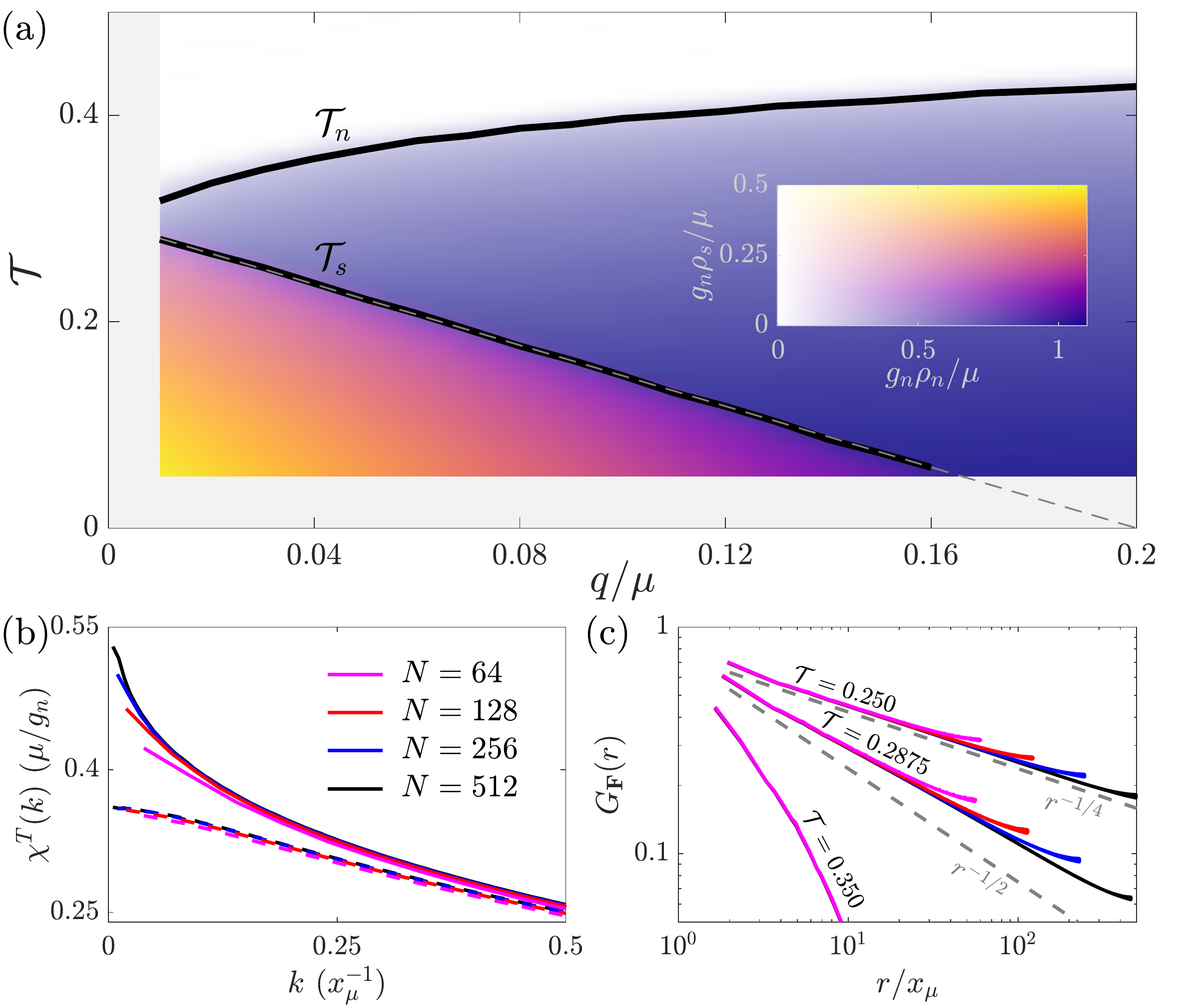}
\caption{\label{qdependence}(a) $(\mathcal{T},q)$ phase diagram showing mass and spin superfluid transitions (solid black lines). The colormap gives the mass and spin superfluid densities. The spin BKT temperature decreases linearly with increasing $q$, approaching zero as $q\rightarrow q_0$ (grey dashed line is $\propto 1-q/q_0$ with $q_0=2|g_s|\mu/g_n=0.2\mu$). Results for $N=256$. (b) At $q=0$, the mass fluid response $\chi^T(k)$ (solid lines) exhibits a system size dependence. For comparison, $\chi^T(k)$ at $q=0.01\mu$ (dashed lines) converge for increasing $N$. (c) Correlations of total spin, showing a cross-over from exponential to algebraic decay at $\mathcal{T}\approx 0.3$ (system sizes as in (b)). Correlations decay close to $r^{-1/2}$ at the cross-over temperature. A sampling time of $2.5\times 10^{6}\hbar/\mu$ is used for the $N=512$ results in (b) and (c) to avoid autocorrelation.}
\end{figure}

The dependence of $\mathcal{T}_n$ and $\mathcal{T}_s$ on $q/\mu$ is shown in Fig.~\ref{qdependence}(a). The linear dependence of $\mathcal{T}_s$ on $q$ follows from evaluating the zero-point of the free energy $F=E-TS$ of a single PCV, which gives $\mathcal{T}_s\propto 1-q/q_0$~\cite{kosterlitz1972,williamson2019}. Here $S=2k_B\ln L$ and $E=\frac{K}{2}\int_{\xi_s}^L r^{-2}\,d^2\mathbf{r}=\pi K\ln (L/\xi_s)$ are the entropy and energy respectively of a single free (but confined) PCV, with $L$ the system size and $K\approx \hbar^2 (1-q/q_0)\mu /2g_nM$ the spin-wave stiffness. Applying the same free-energy argument to vortices in the $m=0$ component would give $\mathcal{T}_n\propto 1+q/q_0$. While this qualitatively captures the increase in $\mathcal{T}_n$ with $q/q_0$, the linear behaviour holds only for large $q/q_0$. The spin superfluid transition extrapolates to zero at $q=q_0$. At these low temperatures we expect ordering behaviour to be affected by both quantum and thermal fluctuations~\cite{sondhi1997,vojta2003,sachdev2011}.

At $q=0$ the order parameter manifold is $\mathrm{SO}(3)$~\cite{Ho1998a,StamperKurn2013a}, combining the symmetry of the full spin vector $\mathbf{F}=(F_x,F_y,F_z)$ and gauge symmetry into a single manifold, with vortices fundamentally distinct from $U(1)$ systems~\cite{Ho1998a}. The nature of superfluidity and the potential for BKT transitions in $\mathrm{SO}(3)$ systems is not well understood~\cite{kawamura1984,kawamura2010,kawamura1993,wintel1994,wintel1995,kobayashi2019}. We find that the mass fluid response $\chi^T(k)$ exhibits a system size dependence for long wavelengths, see Fig.~\ref{qdependence}(b). This suggests the absence of a $q=0$ mass BKT transition in the thermodynamic limit, consistent with the conclusion of~\cite{kobayashi2019}. Despite this, we still see a cross-over from exponential to algebraic decay in correlations of total spin,
\begin{equation}
G_\mathbf{F}(r)=\frac{\langle\mathbf{F}(\mathbf{0})\cdot \mathbf{F}(\mathbf{r})\rangle}{\langle\mathbf{F}(\mathbf{0})\cdot \mathbf{F}(\mathbf{0})\rangle},
\end{equation}
see Fig.~\ref{qdependence}(c). The cross-over temperature $\mathcal{T}\approx 0.3$ slowly decreases with increasing system-size, and hence may go to zero in the thermodynamic limit. In the finite-sized systems explored here, $G_{\mathbf{F}}(r)$ decays close to $r^{-1/2}$ at the cross-over temperature, rather than as $r^{-1/4}$ typical for $\mathrm{U}(1)$ systems. This is similar to the scaling of low-temperature correlations in a finite-sized ferromagnetic Heisenberg model in 2D~\cite{kapikranian2007}. Importantly, though, the order parameter manifold of the Heisenberg model is $S^2$, not $\mathrm{SO}(3)$, and hence does not support vortices~\cite{toda1962}. A cross-over in the behaviour of correlations around a well-defined temperature has been identified in a Heisenberg antiferromagnet on a triangular lattice, which also has an $\mathrm{SO}(3)$ order parameter manifold~\cite{wintel1994,wintel1995,popov2017}.

\paragraph{Discussion.}

In this work we identified two BKT transitions in the easy-plane phase of a ferromagnetic spin-1 Bose gas. We characterised the transitions in terms of relevant vortices and inter- and intra-component fluid responses, and identified possible deconfinement of PCVs above the spin-superfluid transition. At $q=0$ the fluid response exhibits a system size dependence, however correlations of total spin still exhibit a cross-over from exponential to algebraic decay. It would be interesting to analyse this behaviour for vanishingly small $q$ and how this depends on system size. Extending our work, one could explore the role of non-zero axial magnetization, which adjusts the relative density of the $m = \pm 1$ spin components and modifies the nature of the PCVs~\cite{williamson2021}. One could also explore the role of transverse trapping~\cite{keepfer2022}, which may affect the mass and spin superfluidities differently. 

Our work opens up the possibility to explore temperature quenches across the spin BKT transition and to study nonequilibrium processes such as Kibble-Zurek scaling and coarsening dynamics. Comparisons with the extensive work on zero-temperature quenches to the easy-plane phase~\cite{Sadler2006a,Leslie2009a,Saito2005a,Barnett2011,Saito2007b,Lamacraft2007a,Damski2007a,anquez2016,Williamson2016a,Williamson2016b,williamson2019,prufer2018}, particularly close to $q_0$, could illuminate the role of quantum versus thermal fluctuations in the symmetry breaking and nonequilibrium dynamics~\cite{sondhi1997,vojta2003,sachdev2011}. We have shown that the $m = \pm 1$ spin components decohere for increasing temperature, indicating states beyond the low-temperature $\mathrm{U}(1)\times \mathrm{SO}(2)$ manifold. For very high temperatures we expect restoration to the full $\mathrm{SU}(3)$ manifold; our work paves the way to explore how such a phase emerges.

 %We find that $\mathcal{T}_n$ increases as $q/q_0$ increases, approaching the value $\mathcal{T}_n\approx 0.42$ at $q\approx q_0$, agreeing with values obtained for a scalar condensate~\cite{addIn}????. This is to be expected, as for $q\approx q_0$, the mass density is almost entirely in the $m=0$ component and hence we expect the mass superfluidity to behave the same as a scalar condensate.

\paragraph{Acknowledgements.} A.J.G. acknowledges Tom Billam for useful discussions. This research was supported by the Australian Research
Council Centre of Excellence for Engineered Quantum Systems (Project No. CE170100009) and the Australian Research Council
Centre of Excellence in Future Low-Energy Electronics Technologies (Project No. CE170100039). P.B.B acknowledges support from the Marsden Fund of the Royal Society of New Zealand. X.Y. acknowledges support from the National Natural Science Foundation of China (Grant No. 12175215), the National Key Research and Development Program of China (Grant No. 2022YFA 1405300) and  NSAF (Grant No. U1930403). We acknowledge the use of New Zealand eScience Infrastructure (NeSI) high performance computing facilities.

%This will be sensitive to the spin interaction strength, which could be explored in experiments using different species~\cite{kempen2002,Chang2005a,widera2006,huh2020} or potentially via Feshbach resonances~\cite{hamley2009,zhang2009,papoular2010}.

%The energetics of PCVs are notably different from scalar vortices, due to an internal ``stretch'' energy~\cite{Turner2009,Williamson2016c} arising from the composite nature of the vortices and an associated damping due to coupling to spin waves~\cite{williamson2021}. This will likely play an important role in any non-equilibrium dynamics across the spin BKT transition. It is also worth investigating whether this affects the statistical behaviour of bound PCVs below the spin BKT transition.

%We have focussed on the case of weak spin interaction relevant to $^{87}$Rb. It would be interesting to explore the role of interaction strength on the transition, which may be possible using Feshbach resonances~\cite{hamley2009,zhang2009,papoular2010} or by comparing different species~\cite{huh2020}. For example, one could explore how the spin BKT transition emerges from the non-interacting limit (c.f.~\cite{fletcher2015}), where the three spin components behave as independent superfluids. 

%apsrev4-2.bst 2019-01-14 (MD) hand-edited version of apsrev4-1.bst
%Control: key (0)
%Control: author (8) initials jnrlst
%Control: editor formatted (1) identically to author
%Control: production of article title (0) allowed
%Control: page (0) single
%Control: year (1) truncated
%Control: production of eprint (0) enabled
%

\end{document}